\documentclass[conference]{IEEEtran}
\IEEEoverridecommandlockouts

\usepackage{cite}
\usepackage{amsmath,amssymb,amsfonts}
\usepackage{algorithmic}
\usepackage{graphicx}
\usepackage{textcomp}
\usepackage[table]{xcolor}
\usepackage{multirow}
\usepackage{booktabs}
\usepackage{array}
\usepackage{float}
\usepackage{comment}
\usepackage[hidelinks]{hyperref}

\def\BibTeX{{\rm B\kern-.05em{\sc i\kern-.025em b}\kern-.08em
    T\kern-.1667em\lower.7ex\hbox{E}\kern-.125emX}}
\newcommand{\blue}[1]{\textcolor{black}{#1}}
\newcommand{\red}[1]{{\color{black}#1}}
\definecolor{rebuttalpurple}{RGB}{0,0,0}
\newcommand{\rev}[1]{\textcolor{rebuttalpurple}{#1}}
\newcommand{\keyphrase}[1]{\noindent\textbf{#1}}
\begin{document}

\title{Phonemes vs. Projectors: An Investigation of Speech-Language Interfaces for LLM-based ASR
\thanks{\textsuperscript{*} Corresponding author (ozj@tsinghua.edu.cn).}
\thanks{\textsuperscript{\dag} Work done during internship at TasiTech.}
}

\author{
    \IEEEauthorblockN{Ziwei Li$^1$, Lukuang Dong$^2$, Saierdaer Yusuyin$^3$\textsuperscript{\dag}, Xianyu Zhao$^2$, Zhijian Ou$^{1,2}$\textsuperscript{*}}
    \IEEEauthorblockA{$^1$ Speech Processing and Machine Intelligence (SPMI) Lab, Tsinghua University, China}
    \IEEEauthorblockA{$^2$ TasiTech Co., Ltd., China}
    \IEEEauthorblockA{$^3$ School of Computer Science and Technology, Xinjiang University, China}
}


\maketitle

\begin{abstract}
Integrating pretrained speech encoders with large language models (LLMs) is promising for ASR, but performance and data efficiency depend on the speech-language interface. \blue{A common choice is a learned projector that maps encoder features into the LLM embedding space, whereas an alternative is to expose discrete phoneme sequences to the LLM. Using the same encoder and LLM backbones, we compare phoneme-based and vanilla projector-based interfaces in high-resource English and low-resource Tatar. We also propose a BPE-phoneme interface that groups frequent local phoneme patterns while preserving explicit word-boundary cues for phoneme-to-grapheme generation. On LibriSpeech, the phoneme-based interface is competitive with the vanilla projector, and the BPE-phoneme interface yields further gains. On Tatar, the phoneme-based interface substantially outperforms the vanilla projector. 
We further find that phoneme supervision yields a phoneme-informed hybrid interface that is stronger than the vanilla projector.}
\end{abstract}

\begin{IEEEkeywords}
large language models, LLM-ASR, speech--language interface
\end{IEEEkeywords}

\section{Introduction}
Auto-regressive large language models (LLMs) have excelled in natural language processing (NLP)~\cite{brown2020gpt3,achiam2023gpt}, showing impressive language modeling and text generation capability.
Exploring the potential of LLMs in automatic speech recognition (ASR) has therefore received increasing interest.
Importantly, ASR is a cross-modality task, which is fundamentally different from NLP: it maps continuous speech signals to discrete text sequences.
To exploit auto-regressive LLMs as decoders, most recent systems introduce a speech encoder to connect speech and language.

Remarkably, the study of speech encoders has been active long before the current LLM wave.
By pre-training over large amounts of speech, these encoders learn transferable acoustic representations and can be viewed as \emph{large acoustic models} (LAMs).
Encoders are developed with different pretraining paradigms, including self-supervised learning (e.g., wav2vec~2.0~\cite{baevski2020wav2vec}, HuBERT~\cite{hsu2021hubert}, and WavLM~\cite{chen2022wavlm}), supervised pretraining with graphemic transcription (e.g., Whisper~\cite{radford2023robust}), and weakly phonetic supervision (e.g., Whistle~\cite{yusuyin2025whistle}).

These motivations are pushing ASR research from monolithic end-to-end models (such as attention-based encoder--decoder systems~\cite{chan2016las}) toward combining LAMs and LLMs, which we refer to as \emph{LLM-based ASR} (LLM-ASR).
A common recipe is to decouple the architecture into a speech encoder and an LLM connected by a speech--language interface.
The encoder is responsible for acoustic perception while the LLM serves as a powerful decoder.
Under this decomposition, the interface becomes a key determinant of both recognition performance and data efficiency.
\blue{Different interfaces have been studied to integrate LLMs with ASR, 
ranging from feeding ASR-generated text to an LLM for error correction~\cite{ma2023can}, to using phoneme sequences as discrete prompts~\cite{ma2025llm}, to mapping continuous speech embeddings into the LLM space through learned projectors~\cite{yu2023connecting,chen2023salm,ma2024embarrassingly,pham2024comprehensive}. At the interface level, these methods can be broadly grouped into two paradigms: \emph{continuous interfaces}, which pass learned speech embeddings to the LLM, and \emph{discrete interfaces}, which expose tokenized intermediate units. Other discrete interfaces use k-means-clustered encoder representations as speech tokens~\cite{wang2024comparative}.}

\blue{Among discrete interfaces, phonemes are especially attractive because they provide a linguistically grounded intermediate representation that an LLM can consume as ordinary text-like tokens. Compared with word-level ASR outputs, phoneme sequences more directly separate pronunciation from orthography, and they retain structure that is partially shared across languages (e.g., through IPA). 
By contrast, projector-based continuous interfaces must bridge two separately pretrained representational spaces. Because this alignment is learned post hoc from paired speech--text data alone, it can be less robust when supervision is limited or mismatched.
This motivates our focus on phoneme-based interfaces as a linguistically grounded discrete alternative, especially in settings where paired supervision is limited.}

Projector-based and phoneme-based interfaces in LLM-ASR have rarely been studied side by side under controlled backbones.
This work presents a systematic study of speech--language interface design for LLM-ASR.
We compare projector-based and phoneme-based interfaces in both high-resource and low-resource settings.
\blue{We further propose a \emph{BPE-phoneme} interface that merges frequent local phoneme patterns while retaining explicit word boundaries for the LLM decoder.}

\begin{figure*}[ht]
  \centering
  \includegraphics[width=0.7\textwidth]{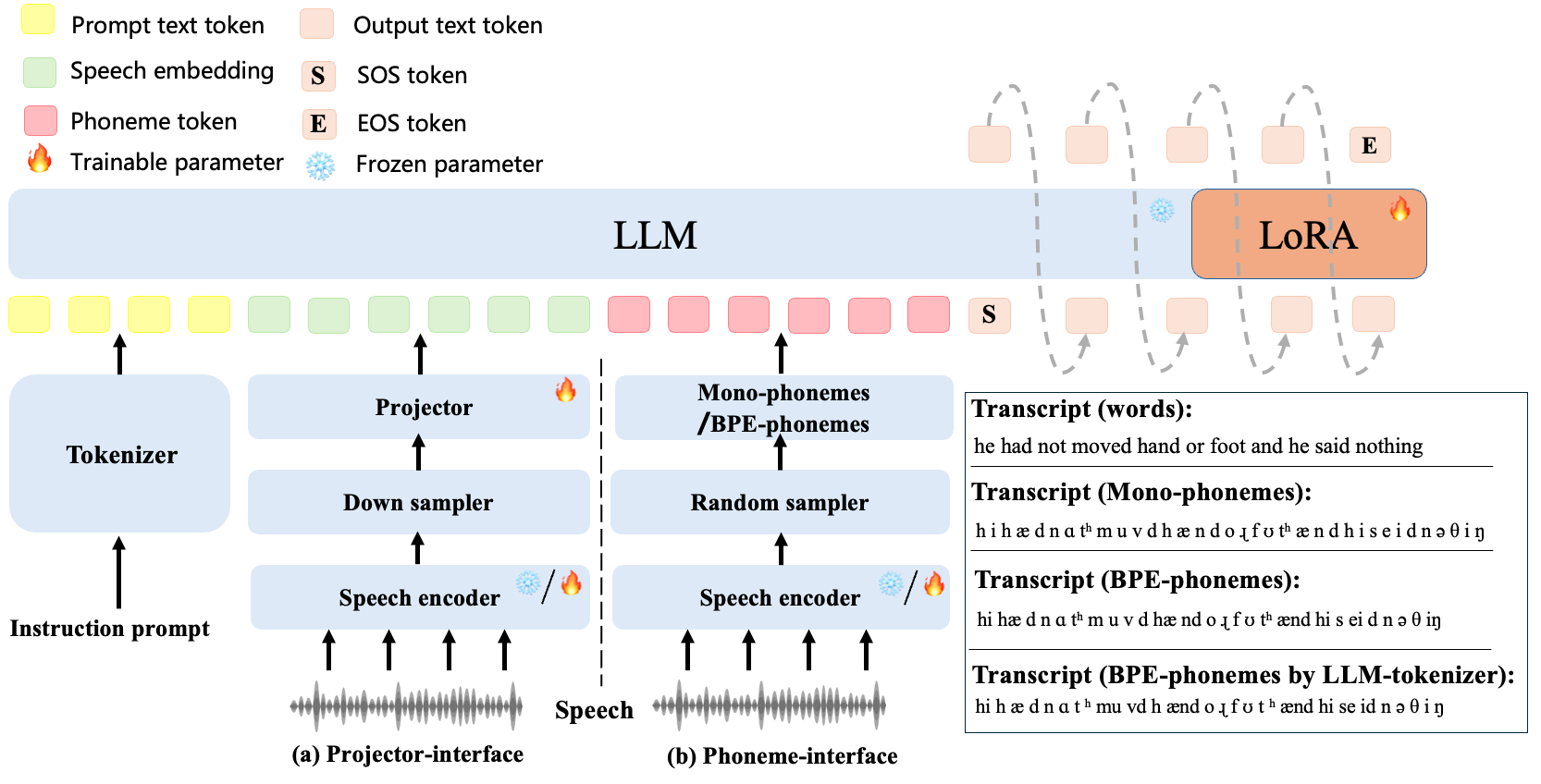}
  \caption{Overview of the two types of speech--language interfaces studied in this work.
  (a) \textbf{Projector-based:} a trainable projector transforms downsampled speech-encoder ouput vectors into the LLM embedding space, followed by LoRA adaptation of the LLM.
  (b) \textbf{Phoneme-based:} we first train a CTC speech-to-phoneme (S2P) model, then feed sampled phoneme sequences into the LLM and adapt it with LoRA for phoneme-to-grapheme (P2G) generation.}
  \label{fig:projectorVSphoneme}
\end{figure*}

Remarkably, the two approaches are not strictly orthogonal.
For instance, a speech encoder can first be fine-tuned with phoneme supervision and then reused in a projector pathway.
\blue{We refer to such systems as \emph{phoneme-informed hybrid interfaces}, and we find that they yield substantially stronger performance.}
Overall, our experiments show that projector-based and phoneme-based methods are competitive across model configurations on LibriSpeech (960h), \blue{while the best-performing configuration in our study is achieved by the BPE-phoneme interface.}
\blue{On low-resource Tatar (20h), the phoneme-based interface substantially outperforms the projector-based interface. Both systems use the same frozen pretrained Whistle-large encoder, so the improvement can be attributed to the interface rather than extra target-data encoder adaptation.}

\rev{Thus, the contribution of this paper is a controlled interface-level study,
not a claim that phoneme interfaces dominate all projector or speech-unit alternatives.
The key findings are BPE-phoneme
prompting, a stronger
phoneme-informed hybrid projector, and the low-resource advantage on Tatar.}

\section{Related work}

\subsection{End-to-End and Pretrained ASR}
Modern end-to-end ASR is commonly formulated with CTC~\cite{graves2006connectionist}, RNN-T~\cite{graves2012sequence}, or attention-based encoder--decoder models~\cite{chorowski2014end}, which unify acoustic modeling, alignment, and decoding in a single neural network.
Large-scale speech pretraining further improves data efficiency: self-supervised approaches such as wav2vec~\cite{baevski2020wav2vec}, HuBERT~\cite{hsu2021hubert}, and WavLM~\cite{chen2022wavlm} learn transferable acoustic representations from unlabeled audio, while supervised pretraining such as Whisper~\cite{radford2023robust} and Whistle~\cite{yusuyin2025whistle} leverages massive weakly labeled multilingual data.

\subsection{LLM-based ASR}
LLM-based ASR aims to leverage the language modeling capacity and world knowledge of LLMs~\cite{achiam2023gpt,touvron2023llama,bai2023qwen} for speech recognition.
Recent systems reformulate ASR as conditional generation with a pretrained speech encoder and a pretrained LLM decoder, e.g., Qwen3-ASR~\cite{shi2026qwen3}, OLMoASR~\cite{ngo2025olmoasr}, and Omnilingual-ASR~\cite{omnilingual2025omnilingual}.
\blue{Complementary lines of work bridge continuous speech features and LLM representations with learned connectors, adapters, or projectors, either with the LLM frozen or with lightweight PEFT such as LoRA~\cite{yu2023connecting,chen2023salm,ma2024embarrassingly,pham2024comprehensive}. For LibriSpeech 960h, representative recent baselines include Q-Former connectors~\cite{yu2023connecting} and projector/adaptor-based systems~\cite{chen2023salm,ma2024embarrassingly,pham2024comprehensive}. We later compare our best model to these systems to contextualize the absolute WERs reported in this paper.}
Beyond ASR-focused systems, speech-capable multimodal LLMs such as Gemini~\cite{team2023gemini} and AudioPaLM~\cite{rubenstein2023audiopalm} further emphasize that the speech--LLM bridge is a recurring bottleneck.

\subsection{Speech--Language Interfaces and Phoneme Modeling}
A central design choice in LLM-ASR is the speech--language interface that bridges continuous speech representations and the discrete token space of an LLM.
\emph{Continuous interfaces} learn this bridge with projectors/adapters that map speech-encoder states to LLM embeddings, but may require substantial paired data to learn a robust cross-modal alignment.
\emph{Discrete interfaces} avoid continuous alignment by providing the LLM with discrete sequences, e.g., ASR-generated text for error correction or rescoring~\cite{ma2023can} or pronunciation-level units such as phonemes~\cite{ma2025llm}. 
Pure text corresponds to an ASR-transcript cascade, whereas phonemes retain pronunciation-level uncertainty before orthographic decoding. 
Learned speech units are complementary; prior controlled results found simple k-means units less competitive than continuous features in the studied setting~\cite{wang2024comparative}. 
Phoneme sequences are linguistically grounded and partially shared across languages through IPA~\cite{international1999handbook}, which makes them appealing for low-resource and cross-lingual transfer.
Subword modeling such as BPE~\cite{sennrich2016neural} can also be applied to phoneme sequences to form compact multi-phoneme units that capture frequent phonotactic patterns. \blue{Applying BPE to phoneme sequences has been explored in end-to-end ASR~\cite{wang2020investigation,zeineldeen2021systematic,yusuyin2023investigation}. Our setting is different: we use BPE-phoneme units not as the final labels of a standalone end-to-end recognizer, but as an intermediate discrete interface between an S2P model and an LLM-based P2G model.}
While phoneme supervision has shown promise in multilingual ASR pretraining (e.g., Whistle~\cite{yusuyin2025whistle}), controlled comparisons of phoneme-based and projector-based interfaces within the same LLM-ASR backbone remain limited; this work aims to fill that gap. \rev{Our phoneme pathway builds on the S2P--P2G LLM-ASR framework~\cite{ma2025llm} and its sampling-marginalization extension~\cite{ma2025phoneme}, but uses them as one side of a controlled interface comparison and extends the interface with BPE-phoneme and phoneme-informed projector variants.}

\section{Method}

\subsection{Projector-based Interface}

As shown in Fig. \ref{fig:projectorVSphoneme}(a), the projector-interface approach consists of a speech encoder, a trainable projector, and an auto-regressive LLM decoder. 
For each utterance, we denote the speech signal as $\mathbf{x}$ and the target transcript as $\mathbf{y}$.
The speech encoder extracts frame-level acoustic representations:
\begin{equation}
\mathbf{H}_\mathbf{x} = \mathrm{Enc}(\mathbf{x}),
\end{equation}
where $\mathbf{H}_\mathbf{x} \in \mathbb{R}^{T \times d_s}$, 
$T$ is the number of acoustic frames and $d_s$ is the encoder feature dimension.
As the speech feature sequence is often very long for the LLM to tackle, it is usual to downsample the speech features with a downsampler:
\begin{equation}
\tilde{\mathbf{H}}_\mathbf{x}
=
\mathrm{DownSampling}_k(\mathbf{H}_\mathbf{x}),
\end{equation}
where $\tilde{\mathbf{H}}_\mathbf{x} \in \mathbb{R}^{N \times k d_s}$. Every $k$ consecutive frames in the feature dimension are concatenated to perform $k$ times downsampling and $N = \lfloor T/k \rfloor$.
The downsampled acoustic features are then projected into the LLM's input embedding space.
In our experiments, we use a hidden layer followed by a ReLU activation and a regression layer as the projector, denoted as:

\begin{equation}
\mathbf{E}_\mathbf{x}
=
\mathrm{Linear}_2
\big(
\mathrm{ReLU}
(
\mathrm{Linear}_1(\tilde{\mathbf{H}}_\mathbf{x})
)
\big),
\end{equation}
where $\mathrm{Linear}_1: \mathbb{R}^{k d_s} \rightarrow \mathbb{R}^{d_h}$ maps acoustic features to the hidden space, and $\mathrm{Linear}_2: \mathbb{R}^{d_h} \rightarrow \mathbb{R}^{d}$ maps hidden representations to the LLM embedding space.
Here $d_h$ denotes the hidden dimension of the projector and 
$d$ is the embedding dimension used by LLM.
The resulting $\mathbf{E}_x \in \mathbb{R}^{N \times d}$ serves as the speech-conditioned ``prefix'' embeddings for the LLM to generate the transcript.
In our experiments we evaluate both \emph{vanilla} projector training with a frozen pretrained encoder, and a \blue{\emph{phoneme-informed projector}} variant that initializes the encoder with phoneme fine-tuning (Table~\ref{tab:librispeech_interface_compare}).

\red{The projector pathway is optimized in two phases, following the common recipe for LLM-based ASR~\cite{ma2024embarrassingly}. In Pr-Stage1, the speech encoder and LLM are frozen and only the projector is trained to align downsampled acoustic features with the LLM embedding space. In Pr-Stage2, the learned projector is kept fixed while the LLM is adapted with LoRA, enabling the decoder to better use the speech-conditioned prefix.}

\subsection{Phoneme-based Interface}

As illustrated in Fig. \ref{fig:projectorVSphoneme}(b), 
\rev{following prior LLM-based S2P--P2G ASR frameworks~\cite{ma2025llm,ma2025phoneme},} the discrete-interface approach factorizes ASR into a speech-to-phoneme (S2P) model and a phoneme-to-grapheme (P2G) model.
Given an input speech feature sequence $\mathbf{x} = (x_1, x_2,\dots, x_T)$, the objective is to predict the most probable grapheme sequence $\mathbf{y} = (y_1, y_2,\dots, y_L)$. 
We marginalize over all possible intermediate phoneme sequences $\mathbf{h}$: 
\begin{equation}
p(\mathbf{y}\mid \mathbf{x}) = \sum_{\mathbf{h}} p(\mathbf{h}\mid \mathbf{x})\, p(\mathbf{y}\mid \mathbf{h}),
\end{equation}
where $p(\mathbf{h}\mid\mathbf{x})$ is implemented by a CTC-based S2P model on top of a speech encoder, and $p(\mathbf{y}\mid\mathbf{h})$ is implemented by an auto-regressive LLM fine-tuned for P2G generation.
The training in the discrete-interface approach consists of two stages.

\keyphrase{Ph-Stage 1: S2P with CTC.}
We add a CTC~\cite{graves2006connectionist} classifier over an IPA-based phoneme vocabulary on top of the speech encoder and optimize the standard CTC objective to obtain $p(\mathbf{\pi}\mid \mathbf{x})$ over alignment paths $\mathbf{\pi}$.

\keyphrase{Ph-Stage 2: P2G with Simplified Sampling-$K$ Marginalization (S-SKM).}
To expose phoneme uncertainty to the LLM during training, we draw $K$ CTC path samples $\mathbf{\mathbf{\pi}}^{(k)}\sim p(\mathbf{\pi}\mid \mathbf{x})$, collapse repeats and blanks with the CTC operator $\mathcal{B}(\cdot)$, and obtain phoneme sequences $\mathbf{h}^{(k)}=\mathcal{B}(\mathbf{\pi}^{(k)})$.
Here, $K$ denotes the number of sampled phoneme sequences used to approximate the marginalization over phoneme uncertainty. We then train the LLM with a Monte Carlo approximation~\cite{ma2025phoneme}.
\begin{equation}
\mathcal{L}_{\text{S-SKM}}
=
-
\log
\left(
\frac{1}{K}
\sum_{k=1}^{K}
p(\mathbf{y} \mid \mathbf{h}^{(k)})
\right),
\end{equation}
where $K$ is the number of samples (we use $K{=}8$).

\red{
\keyphrase{Decoding.}
At inference time, we use Top-$K$ Marginalization (TKM) for rescoring rather than relying on a single phoneme sequence~\cite{ma2025llm}. The S2P beam search first returns the top-$K$ phoneme hypotheses $\{\mathbf{h}^{(k)}\}_{k=1}^{K}$ with their beam scores. The P2G model then produces the top-$S$ text hypotheses for each phoneme hypothesis, and the union of these hypotheses is deduplicated into a candidate set $\mathcal{Y}$. The final transcript is selected by
\begin{equation}
\label{eq:tkm}
\hat{\mathbf{y}} = \arg\max_{\mathbf{y}\in\mathcal{Y}}
\sum_{k=1}^{K} p(\mathbf{h}^{(k)}\mid\mathbf{x})\, p(\mathbf{y}\mid\mathbf{h}^{(k)}).
\end{equation}
When $\mathbf{y}$ is not generated among the top-$S$ candidates conditioned on $\mathbf{h}^{(k)}$, its conditional probability under that phoneme hypothesis is treated as zero in Eq.~(\ref{eq:tkm}).
}

\subsection{BPE-Phoneme Modeling}
\blue{Mono-phoneme prompts expose the LLM to atomic pronunciation units, but they do not contain explicit word-boundary information. We therefore propose a \emph{BPE-phoneme} interface that merges frequent local phoneme patterns while preserving whitespace between phoneme words. This design is motivated by prior phone-based subword work~\cite{wang2020investigation,zeineldeen2021systematic,yusuyin2023investigation}, but here it is used as the discrete interface between S2P and the LLM. In our implementation, BPE-phoneme is not a prompt-shortening trick: because it retains word boundaries, it can even produce slightly longer LLM token sequences, as quantified later in our analysis. We therefore interpret its gain as coming from boundary-aware segmentation and chunked local phonotactic units rather than fewer LLM input tokens. We learn BPE merges on the phoneme transcripts from the training set using SentencePiece~\cite{kudo2018sentencepiece} and use the resulting BPE-phoneme units as CTC targets in Stage~1. 
The BPE-phoneme sequences generated from the S2P CTC model carry whitespaces, which are converted to phoneme words and fed to the LLM in Stage~2.}

\section{Experimental Setup}

\subsection{Datasets}
We evaluate interface designs in both high-resource and low-resource settings. \rev{These two settings are intended to isolate interface effects in a data-rich English condition and a label-scarce target-language condition, rather than to serve as an exhaustive multilingual benchmark.}
\textbf{English (high-resource):} We use LibriSpeech (960h)~\cite{panayotov2015librispeech}, validate on \textit{dev-other}, and report WER on \textit{test-clean} and \textit{test-other}.
\textbf{Tatar (low-resource):} We use the Tatar (\texttt{tt}) subset of Common Voice~\cite{ardila2020commonvoice} v11.0 (\textbf{CV-tt}, 20h). We validate on \textit{dev} and report WER on \textit{dev} and \textit{test}.

\subsection{Model Backbones}
\noindent\blue{\textbf{Speech encoders.} On LibriSpeech we evaluate WavLM-base and WavLM-large (self-supervised pretraining) as well as Whistle-small (phoneme-supervised pretraining) to cover diverse encoder pretraining paradigms. Whistle-small and Whistle-large are multilingual phoneme-based speech encoders pretrained on 10 Common Voice languages, including Tatar~\cite{yusuyin2025whistle}. For CV-tt we use the same pretrained frozen Whistle-large encoder for both projector-based and phoneme-based systems. We do not fine-tune the encoder on the 20h Tatar data in either case, yielding a fair low-resource comparison that isolates interface effects from extra target-task encoder adaptation.}

\keyphrase{LLMs.} We use Qwen3-Base as the LLM backbone. Unless otherwise specified we adapt Qwen3-1.7B; we also report scaling results with Qwen3-8B in Table~\ref{tab:phoneme-interface-ablation}.
\blue{We inject LoRA~\cite{hu2022lora} into both the self-attention (\texttt{q/k/v/o}) and MLP (\texttt{gate/up/down}) modules with rank $r{=}512$ and scaling $\alpha{=}256$, resulting in about \textit{0.5B} and \textit{1.4B} trainable parameters for \textit{Qwen3-1.7B-Base} and \textit{Qwen3-8B-Base}, respectively.}

\subsection{Pronunciation Lexicons and Phoneme Targets}
We lowercase transcripts and remove pronunciation-irrelevant punctuation.
For LibriSpeech, we build an IPA lexicon by converting CMUdict pronunciations and training Phonetisaurus~\cite{novak2016phonetisaurus} to expand coverage; for CV-tt we use Phonetisaurus with pretrained LanguageNet \rev{grapheme-to-phoneme (G2P)} models~\cite{hasegawa2020grapheme} to generate IPA pronunciations for in-vocabulary words. 
\rev{These phoneme targets are derived from transcripts and impose a pronunciation-level structure rather than copying text tokens; Whistle reports that weak G2P-derived phonetic supervision can outperform subword/text supervision when transcripts are available~\cite{yusuyin2025whistle}.}
Mono-phoneme experiments use the resulting IPA sequences as CTC targets.
For BPE-phoneme, we train a SentencePiece~\cite{kudo2018sentencepiece} BPE tokenizer on phoneme sequences (keeping word boundaries) and use the BPE-phoneme units as CTC targets.

\subsection{Speech--Language Interfaces}

\keyphrase{\blue{Projector-based (vanilla and phoneme-informed).}}
We downsample encoder states with rate $k{=}5$ and map them to the LLM embedding size using a two-layer MLP projector ($d_h{=}2048$, about 20M parameters).
We evaluate (i) a {vanilla projector} with a frozen pretrained encoder and (ii) a \blue{{phoneme-informed projector}} that initializes the encoder from Ph-Stage1 phoneme fine-tuning (mono-phoneme or BPE-phoneme).

\keyphrase{Phoneme-based.}
We use an IPA phoneme vocabulary for CTC-based S2P modeling.
For mono-phoneme, the S2P output is a sequence of mono-phonemes and thus contains no word-boundary information; before feeding it into the LLM tokenizer, we remove all whitespace between adjacent mono-phoneme tokens to form a continuous phoneme string.
For BPE-phoneme, we first merge BPE-phoneme units back into word-level phoneme sequences (phoneme words), and then pass the resulting sequence to the LLM tokenizer for tokenization. \blue{The resulting average utterance lengths in terms of LLM tokens are discussed later in the analysis.}

\subsection{Training}
\keyphrase{Projector-based.}
\emph{Pr-Stage1}: freeze the speech encoder and LLM; train only the projector.
\emph{Pr-Stage2}: freeze the projector; adapt the LLM with LoRA.
For the \blue{{phoneme-informed projector}}, we initialize the encoder from \emph{Ph-Stage1} and then run Pr-Stage1/2.

\keyphrase{Phoneme-based.}
\emph{Ph-Stage1}: fine-tune the speech encoder for phoneme prediction with CTC (mono-phoneme or BPE-phoneme targets).
\emph{Ph-Stage2}: sample $K{=}8$ phoneme sequences per utterance from the frozen S2P model and fine-tune the LLM for P2G generation with S-SKM and LoRA.
\blue{For CV-tt, we modify the protocol to match the low-resource setting: both the projector-based and phoneme-based systems use the same pretrained Whistle-large encoder and keep it frozen throughout training. No Tatar-specific encoder fine-tuning is applied in either case; the only difference between the two systems is the interface. We therefore do not introduce an additional phoneme-informed projector baseline on CV-tt: once the encoder is fixed and frozen, the comparison is directly between continuous and discrete interfaces.}

\begin{table}[t]
\centering
\caption{Ph-Stage1 optimization settings for phoneme-based interfaces.}
\vspace{-0.2cm}
\label{tab:ph_stage1_optimization}
\scriptsize
\setlength{\tabcolsep}{2pt}
\renewcommand{\arraystretch}{1.12}
\begin{tabular}{@{}>{\raggedright\arraybackslash}m{0.17\linewidth}>{\raggedright\arraybackslash}m{0.19\linewidth}>{\centering\arraybackslash}m{0.09\linewidth}>{\raggedright\arraybackslash}m{0.31\linewidth}>{\raggedleft\arraybackslash}m{0.10\linewidth}@{}}
\toprule
\textbf{Target} & \textbf{Encoder} & \textbf{Batch} & \textbf{LR schedule} & \textbf{Steps} \\
\midrule
Mono-phoneme & \shortstack[l]{WavLM-Base,\\Whistle-Small} & 300 & Peak $2.5\times10^{-4}$; 25k warmup & 200k \\
\midrule
BPE-phoneme & WavLM-Base & 300 & Peak $6\times10^{-5}$; 20k warmup & 200k \\
\midrule
BPE-phoneme & WavLM-Large & 120 & Start $3\times10^{-5}$; 10\% up, 40\% hold, then decay & 320k \\
\bottomrule
\end{tabular}
\vspace{0.1cm}
\end{table}

\subsection{\red{Additional Optimization Details}}
\label{sec:opt_details}

\begin{table*}[!t]
\caption{Comparison of projector-based and phoneme-based interfaces on LibriSpeech (LLM: Qwen3-1.7B-Base + LoRA). E1--E2 are \emph{vanilla projector} baselines with frozen pretrained encoders. \blue{E3--E5 use phoneme-finetuned (FT) encoders (Ph-Stage1), enabling a direct comparison between the {phoneme-informed projector} (Proj.) and the phoneme-based interface (Phon.).} PER is reported only for phoneme-finetuned encoders; (--) indicates not applicable. \blue{For the BPE-phoneme interface we use BPE size 100, selected by the ablation in Table~\ref{tab:bpe_size_ablation}.}}
\label{tab:librispeech_interface_compare}
\centering
\scriptsize
\setlength{\tabcolsep}{4pt}
\renewcommand{\arraystretch}{1.06}
\begin{tabular}{l l c cc cc cc}
\toprule
\multirow{3}{*}{\textbf{ID}} & \multirow{3}{*}{\textbf{Speech Encoder}} & \multirow{3}{*}{\shortstack{\textbf{Encoder}\\\textbf{Params}}}
& \multicolumn{2}{c}{\textbf{PER($\downarrow$)}}
& \multicolumn{4}{c}{\textbf{WER($\downarrow$)}} \\
\cmidrule(lr){4-5}\cmidrule(lr){6-9}
& & 
& \multirow{2}{*}{\textbf{clean}} & \multirow{2}{*}{\textbf{other}}
& \multicolumn{2}{c}{\textbf{clean}} 
& \multicolumn{2}{c}{\textbf{other}} \\
\cmidrule(lr){6-7}\cmidrule(lr){8-9}
& & 
&  &  
& Proj. & Phon. & Proj. & Phon. \\
\midrule
E1 & Whistle-small                           & 90M   & --   & --   & 5.94 & --   & 12.66 & --   \\
E2 & WavLM-base                              & 94.7M & --   & --   & 3.08 & --  & 7.66  & --  \\
\midrule
E3 & Whistle-small + Mono-phoneme-FT         & 90M   & 1.78 & 4.73 & 4.11 & 4.28 & 9.18  & 9.46 \\
E4 & WavLM-base + Mono-phoneme-FT            & 94.7M & 1.10 & 3.21 & 3.03 & 3.37 & 7.02  & 7.56 \\
E5 & WavLM-base + BPE-phoneme-FT             & 94.7M & \textbf{1.06} & \textbf{3.29} & 2.85 & \textbf{2.78} & 6.79 & \textbf{6.75} \\
\bottomrule
\end{tabular}
\end{table*}

\red{
\keyphrase{Phoneme-based interface.}
For the phoneme-based interface, Ph-Stage1 uses encoder- and target-specific optimization settings, as summarized in Table~\ref{tab:ph_stage1_optimization}.
Mono-phoneme models based on WavLM-Base and Whistle-Small share the same schedule, whereas BPE-phoneme models use lower or staged learning-rate schedules depending on the encoder.
For the phoneme-to-grapheme stage, all models are trained using the same optimization strategy. We employ a linear warmup followed by cosine annealing, with a peak learning rate of $2\times10^{-4}$ and a minimum learning rate of $1\times10^{-5}$. The batch size is 80, training is conducted for 2 epochs, and the ratio of warmup steps to cosine-decay steps is set to $1{:}10$.

\keyphrase{Projector-based interface.}
\rev{For the projector-based interface, we follow~\cite{ma2024embarrassingly}: Pr-Stage1 uses batch size 80, 1000-step warmup to $1\times10^{-4}$, and early stopping after 5 stagnant validations; Pr-Stage2 uses batch size 128, fixed $1\times10^{-5}$ learning rate, and the same early-stopping criterion.}
}

\begin{table}[t]
\caption{\blue{BPE-size ablation for the BPE-phoneme S2P model on LibriSpeech using WavLM-base. BPE size 100 works best and is used in all subsequent BPE-phoneme experiments.}}
\vspace{-0.2cm}
\label{tab:bpe_size_ablation}
\centering
\scriptsize
\setlength{\tabcolsep}{6pt}
\renewcommand{\arraystretch}{1.05}
\color{black}
\begin{tabular}{lccc}
\toprule
\textbf{BPE size} & \textbf{PER clean} & \textbf{PER other} & \textbf{S2P Params (M)} \\
\midrule
100 & \textbf{1.06} & \textbf{3.29} & 94.46 \\
200 & 1.23 & 3.54 & 94.54 \\
1000 & 1.48 & 4.54 & 95.10 \\
\bottomrule
\end{tabular}
\vspace{0.1cm}
\end{table}

\section{Results}

\subsection{High-Resource Evaluation on LibriSpeech}
\label{sec:results_librispeech}
Table~\ref{tab:librispeech_interface_compare} reports LibriSpeech results under controlled backbones (fixed LLM family and comparable LoRA adaptation).
\textbf{Vanilla projector baselines (E1--E2) are weak}, suggesting that learning a robust continuous speech--text alignment from paired data alone can be challenging.
{\textbf{Phoneme-informed hybrid interfaces are substantially stronger}: after Ph-Stage1 encoder fine-tuning (E3--E5), the phoneme-informed projector improves substantially and becomes competitive with the phoneme-based interface.} \rev{Thus phoneme supervision is useful not only as a discrete interface, but also as an auxiliary structure that makes encoder representations easier for a continuous projector to align with the LLM.}
Under the same phoneme-finetuned encoder, the proposed \textbf{BPE-phoneme} interface yields the best WER (2.78/6.75), slightly outperforming the corresponding phoneme-informed projector (2.85/6.79). 
{We examine the BPE vocabulary size in Table~\ref{tab:bpe_size_ablation}: BPE size 100 gives the best PER on both test subsets, while larger vocabularies (200 and 1000) degrade performance. We therefore use BPE size 100 in all other BPE-phoneme experiments.}

Table~\ref{tab:projector_stages} breaks down the two-stage projector training (Pr-Stage1/2), confirming that (i) LoRA adaptation of the LLM in Pr-Stage2 consistently improves WER and (ii) phoneme fine-tuning yields more projector-friendly representations, especially on \textit{test-other}.

\begin{table}[t]
\caption{Stage-wise WER (\%) on LibriSpeech for the projector interface (Pr-Stage1 vs.\ Pr-Stage2; LLM: Qwen3-1.7B + 0.5B LoRA).}
\vspace{-0.2cm}
\label{tab:projector_stages}
\centering
\scriptsize
\setlength{\tabcolsep}{4pt}
\renewcommand{\arraystretch}{1.05}
\begin{tabular}{l l l cc}
\toprule
\textbf{ID} & \textbf{Speech Encoder} & \textbf{Stage} & \textbf{clean} & \textbf{other} \\
\midrule
E1-Pr1 & \multirow{2}{*}{Whistle-small} & Pr-Stage1 & 7.43 & 14.72 \\
E1-Pr2 &                                & Pr-Stage2 & \textbf{5.94} & \textbf{12.66} \\
\midrule
E2-Pr1 & \multirow{2}{*}{WavLM-base} & Pr-Stage1 & 4.10 & 9.18 \\
E2-Pr2 &                             & Pr-Stage2 & \textbf{3.08} & \textbf{7.66} \\
\midrule
E3-Pr1 & \multirow{2}{*}{Whistle-small + Mono-phoneme-FT} & Pr-Stage1 & 5.38 & 10.81 \\
E3-Pr2 &                                                   & Pr-Stage2 & \textbf{4.11} & \textbf{9.18} \\
\midrule
E4-Pr1 & \multirow{2}{*}{WavLM-base + Mono-phoneme-FT} & Pr-Stage1 & 4.11 & 8.35 \\
E4-Pr2 &                                                & Pr-Stage2 & \textbf{3.03} & \textbf{7.02} \\
\midrule
E5-Pr1 & \multirow{2}{*}{WavLM-base + BPE-phoneme-FT} & Pr-Stage1 & 3.57 & 7.69 \\
E5-Pr2 &                                               & Pr-Stage2 & \textbf{2.85} & \textbf{6.79} \\
\bottomrule
\end{tabular}
\vspace{0.1cm}
\end{table}



\begin{table}[t]
\caption{\blue{Ablation on LibriSpeech for the BPE-phoneme interface} (BPE{=}100): scaling the speech encoder and the LLM (LoRA-adapted).}
\vspace{-0.2cm}
\label{tab:phoneme-interface-ablation}
\centering
\scriptsize
\setlength{\tabcolsep}{3pt}
\renewcommand{\arraystretch}{1.06}

\resizebox{\columnwidth}{!}{%
\begin{tabular}{l l l cc cc}
\toprule
\multirow{2}{*}{\textbf{ID}} &
\multirow{2}{*}{\shortstack{\textbf{Speech Encoder}\\\textbf{(BPE-phoneme-FT)}}} &
\multicolumn{1}{c}{\multirow{2}{*}{\textbf{LLM}}} &
\multicolumn{2}{c}{\textbf{PER} ($\downarrow$)} &
\multicolumn{2}{c}{\textbf{WER} ($\downarrow$)} \\
\cmidrule(lr){4-5}\cmidrule(lr){6-7}
& & & \textbf{clean} & \textbf{other} & \textbf{clean} & \textbf{other} \\
\midrule
E5  & \multirow{2}{*}{WavLM-base}  & Qwen3-1.7B-Base & \multirow{2}{*}{1.06} & \multirow{2}{*}{3.29} & 2.78 & 6.75 \\
E6  &                             & Qwen3-8B-Base   &  &  & 2.43 & 6.00 \\
\midrule
E7  & \multirow{2}{*}{WavLM-large} & Qwen3-1.7B-Base & \multirow{2}{*}{0.70} & \multirow{2}{*}{1.71} & 2.13 & 4.13 \\
E8  &                             & Qwen3-8B-Base   &  &  & \textbf{1.97} & \textbf{3.78} \\
\bottomrule
\end{tabular}
\vspace{0.1cm}
}
\end{table}

\blue{Finally, under the BPE-phoneme interface, we scale either the speech encoder or the LLM. Table~\ref{tab:phoneme-interface-ablation} indicates such scaling improves final WER and that reductions in PER correlate with downstream WER gains.}
In particular, once PER is reduced by encoder scaling, increasing the LLM size yields further WER improvements, reflecting the LLM's role in resolving lexical and contextual ambiguity from phoneme inputs. 
The best overall result in this paper (E8) is obtained by this scaled BPE-phoneme interface.

The BPE-phoneme gain should not be attributed to shorter LLM inputs: its prompt is slightly longer than the monophoneme prompt (125 vs. 113 tokens) because it preserves word boundaries, so the benefit is better explained by explicit segmentation cues and multi-phoneme chunks.

\subsection{\blue{Contextualizing Absolute WERs with Recent LibriSpeech-960h LLM-ASR Baselines}}
\blue{To help judge whether the absolute WERs reported in this work are strong, Table~\ref{tab:recent_librispeech_baselines} compares our best system (E8) with representative recent LLM-ASR baselines on LibriSpeech 960h.} \rev{Although our aim is a controlled interface study rather than a standalone SOTA LibriSpeech recipe, the absolute WERs are competitive.} \blue{The comparison is not perfectly matched in backbones, model scale, or training recipe, but it is still informative. Our E8 result (1.97/3.78) is clearly stronger than early frozen-connector systems such as Q-Former~\cite{yu2023connecting} and SALM~\cite{chen2023salm}, and is competitive with SLAM-ASR~\cite{ma2024embarrassingly}. 
Stronger projector recipes apply LoRA to both the speech encoder and the LLM, and can obtain lower WERs~\cite{pham2024comprehensive}. We therefore view the absolute LibriSpeech results of this work as strong and competitive.}

\begin{table}[t]
\caption{\blue{Contextual comparison with representative recent LLM-ASR baselines on LibriSpeech 960h. The goal is not a perfectly matched comparison---backbones, model scales, and tuning recipes differ---but to gauge whether the absolute WERs reported in this work are competitive. Learnable modules are shown in \textbf{bold}.}}
\vspace{-0.2cm}
\label{tab:recent_librispeech_baselines}
\centering
\footnotesize
\setlength{\tabcolsep}{2.5pt}
\renewcommand{\arraystretch}{1.05}
\color{black}
\resizebox{\columnwidth}{!}{%
\begin{tabular}{>{\centering\arraybackslash}p{0.8cm}>{\centering\arraybackslash}p{0.8cm}p{7.4cm}>{\centering\arraybackslash}p{1.3cm}}
\toprule
\textbf{clean} & \textbf{other} & \textbf{Method} & \textbf{Ref.} \\
\midrule
{1.97} & {3.78} & \textbf{WavLM-large} + BPE-phoneme interface + Qwen3-8B \textbf{LoRA} & E8 \\
2.28 & 5.20 & Whisper-large-v2 + 80-query \textbf{Q-Former connector} (24.5M) + Vicuna-13B & \cite{yu2023connecting} \\
2.30 & 4.80 & NeMo Conformer encoder (110M) + \textbf{adapter} + MegatronLLM-2B \textbf{LoRA} + nucleus sampling & \cite{chen2023salm} \\
2.10 & 4.26 & \textbf{HuBERT-Large} + \textbf{projector} (18.88M) + Vicuna-7B & \cite{ma2024embarrassingly} \\
2.29 & 5.67 & HuBERT-Large + \textbf{adapter} + Vicuna-7B frozen (48M learnable) & \cite{pham2024comprehensive} \\
1.78 & 3.62 & HuBERT-Large \textbf{LoRA} + \textbf{adapter} + Vicuna-7B \textbf{LoRA} (65M learnable) & \cite{pham2024comprehensive} \\
1.85 & 3.77 & HuBERT-Large \textbf{LoRA} + \textbf{adapter} + Vicuna-7B \textbf{LoRA} (337M learnable) & \cite{pham2024comprehensive} \\
\bottomrule
\end{tabular}
}
\vspace{0.1cm}
\end{table}

\subsection{Low-Resource Evaluation on Tatar}
\blue{Table~\ref{tab:tatar_results} reports results on CV-tt (20h) using the same pretrained frozen Whistle-large encoder for both systems. This encoder comes from Whistle's multilingual phoneme-based pretraining over 10 Common Voice languages, and Tatar is one of them~\cite{yusuyin2025whistle}. Importantly, neither system fine-tunes the encoder on the 20h Tatar data, which makes the projector-based and phoneme-based comparison fair. 
In this low-resource setting, the phoneme-based interface nearly halves WER, reducing test WER from 33.58\% (projector) to 17.44\%, which highlights its superior data efficiency under limited paired supervision.} \rev{Since both systems use the same transcripts and frozen Whistle-large encoder, this gain supports the value of the phoneme sequence as a structured intermediate representation beyond simple transcript supervision.}

\begin{table}[t]
\caption{\blue{WER (\%) on low-resource CV-tt (20h) using the same frozen pretrained Whistle-large encoder for both systems.}}
\vspace{-0.2cm}
\label{tab:tatar_results}
\centering
\scriptsize
\renewcommand{\arraystretch}{1.0}
\setlength{\tabcolsep}{6pt}
\begin{tabular}{l l cc}
\toprule
\textbf{ID} & \textbf{Interface} & \textbf{dev} & \textbf{test} \\
\midrule
T1 & Projector-based         & 28.10 & 33.58 \\
T2 & Phoneme-based (Mono)    & \textbf{14.14} & \textbf{17.44} \\
\bottomrule
\end{tabular}
\vspace{0.1cm}
\end{table}

\subsection{\red{Training and inference complexity}}
\label{sec:complexity}
\red{
\rev{For the representative E5 setting on 8$\times$A800 GPUs, the projector pipeline takes 5.14h for Pr-Stage1 (44{,}992 iterations) plus 4.11h for Pr-Stage2 (15{,}620 iterations), i.e., $5.14+4.11=9.25$h. The phoneme system with S-SKM ($K{=}8$) has higher per-iteration cost (2.998s vs. 0.411/0.947s), but uses only 7{,}000 iterations and finishes in 5.83h.}

\rev{Table~\ref{tab:inference_efficiency} compares inference efficiency on one A800 GPU. With $K{=}8$, E5-Ph is slower (RTF around 0.13) but gives the best E5 WER. With $K{=}1$, E5-Ph becomes comparable to projector inference speed (0.090/0.097 RTF on test-clean/test-other, compared with 0.095/0.094 for E5-Pr1) while retaining better WER than E5-Pr1 and approaching E5-Pr2, showing a practical speed--accuracy trade-off controlled by $K$.}
}

\begin{table}[t]
\centering
\caption{\rev{Inference efficiency on one A800 GPU. RTF is the decoding time divided by the total audio duration; lower values indicate faster decoding. For TKM decoding, $K$ denotes the number of S2P phoneme hypotheses, and $beam\_size$ denotes the LLM decoding beam size.}}
\vspace{-0.2cm}
\label{tab:inference_efficiency}
\red{
\scriptsize
\setlength{\tabcolsep}{3pt}
\renewcommand{\arraystretch}{1.0}
\begin{tabular}{lcccc}
\toprule
\textbf{Model} & \textbf{Setting} & \textbf{Test Set} & \textbf{RTF} & \textbf{WER}\\
\midrule

\multirow{2}{*}{E5-Pr1} 
& \multirow{2}{*}{$beam\_size=4$} & test-clean & 0.095 & 3.57\\
&         & test-other & 0.094  & 7.69\\

\midrule

\multirow{2}{*}{E5-Pr2} 
& \multirow{2}{*}{$beam\_size=4$} & test-clean & 0.091 & 2.85\\
&         & test-other & 0.093 & 6.79\\

\midrule

\multirow{2}{*}{E5-Ph} 
& \multirow{2}{*}{$K=8$, $beam\_size=4$} & test-clean & 0.130 & 2.78\\
&               & test-other & 0.134 & 6.75\\

\midrule

\multirow{2}{*}{E5-Ph} 
& \multirow{2}{*}{$K=1$, $beam\_size=4$} & test-clean & 0.090  & 2.94\\
&                         & test-other & 0.097 & 7.19\\
\bottomrule
\end{tabular}
\vspace{0.1cm}
}
\end{table}


\section{Conclusion}

We compared projector-based and phoneme-based speech--language interfaces for LLM-ASR under controlled backbones.
\rev{On LibriSpeech, BPE-phoneme gives the best result in this study and E8 is competitive with recent LibriSpeech-960h LLM-ASR baselines; on low-resource CV-tt, the phoneme interface nearly halves WER with the same frozen Whistle-large encoder. Overall, the paper provides a controlled interface-level comparison with two practical additions: BPE-phoneme prompting in the LLM-ASR setting and phoneme-supervised hybrid projector training.}
\rev{Future work includes pronunciation-lexicon-free optimization~\cite{jsaspg}, unified discrete/continuous bridges, and broader domain and language coverage.}

\clearpage
\section{\texorpdfstring{\textcolor{rebuttalpurple}{Generative AI Use Disclosure}}{Generative AI Use Disclosure}}
\rev{Generative AI tools are used in this work only for language editing, polishing, and formatting of the manuscript. They are not
used to generate any core content, research ideas, experimental
designs, results, or major textual parts of the paper. All scientific contributions, including model design, experiments, analysis, and conclusions, are completed by the authors.}

\bibliographystyle{IEEEtran}

\bibliography{references}

\end{document}